\documentclass[final,3p,times]{elsarticle}

\usepackage{graphicx}

\usepackage{amssymb}
\usepackage{amsmath}

\usepackage{lineno}

\usepackage{enumitem}
\setlist{parsep=0pt,listparindent=\parindent}

\usepackage{verbatim} 
\usepackage{subfig} 
\usepackage{units} 
\usepackage{color} 

\newcommand\T{\rule{0pt}{2.6ex}}
\newcommand\B{\rule[-1.2ex]{0pt}{0pt}}

\journal{Nucl.~Instrum.~Meth.~A}

\begin{document}

\begin{frontmatter}



\title{A compact ultra-clean system for deploying radioactive sources
inside the KamLAND detector}

\author[UCB,LBNL]{T.\,I.~Banks\corref{cor1}} 
\cortext[cor1]{Corresponding author.  Email address: \texttt{tbanks@berkeley.edu}}
\author[UCB,LBNL,IPMU]{S.\,J.~Freedman\fnref{fndeceased}}
\fntext[fndeceased]{Deceased}
\author[LBNL]{J.~Wallig}
\author[LBNL]{N.~Ybarrolaza}
\author[Tohoku]{A.~Gando}
\author[Tohoku]{Y.~Gando}
\author[Tohoku]{H.~Ikeda}
\author[Tohoku,IPMU]{K.~Inoue}
\author[IPMU,Tohoku]{Y.~Kishimoto\fnref{present-KamObs}}
\author[Tohoku,IPMU]{M.~Koga}
\author[Tohoku]{T.~Mitsui}
\author[Tohoku,IPMU]{K.~Nakamura}
\author[Tohoku]{I.~Shimizu}
\author[Tohoku]{J.~Shirai}
\author[Tohoku]{A.~Suzuki}
\author[Tohoku]{Y.~Takemoto\fnref{present-IPMU}}
\author[Tohoku]{K.~Tamae}
\author[Tohoku]{K.~Ueshima}
\author[Tohoku]{H.~Watanabe}
\author[Tohoku]{B.\,D.~Xu}
\author[Tohoku]{H.~Yoshida\fnref{present-OsakaGradSci}}
\author[OsakaGradSci]{S.~Yoshida}
\author[IPMU]{A.~Kozlov}
\author[UA]{C.~Grant\fnref{present-UCD}}
\author[UA]{G.~Keefer\fnref{present-LLNL}}
\author[UA,IPMU]{A.~Piepke}
\author[LBNL]{T.~Bloxham}
\author[LBNL,IPMU]{B.\,K.~Fujikawa}
\author[LBNL]{K.~Han}
\author[UCB]{K.~Ichimura\fnref{present-KamObs}}
\author[UCB,LBNL,IPMU]{H.~Murayama}
\author[UCB,LBNL]{T.~O'Donnell}
\author[UCB,LBNL]{H.\,M.~Steiner}
\author[UCB,LBNL]{L.\,A.~Winslow\fnref{present-UCLA}}
\author[Caltech]{D.\,A.~Dwyer\fnref{present-LBNL}}
\author[Caltech]{R.\,D.~McKeown\fnref{present-JLab}}
\author[Caltech]{C.~Zhang\fnref{present-BNL}}
\author[CSU,IPMU]{B.\,E.~Berger}
\author[Drexel]{C.\,E.~Lane}
\author[Drexel]{J.~Maricic\fnref{present-UHM}}
\author[Drexel]{T.~Miletic\fnref{present-Arcadia}}
\author[UHM]{M.~Batygov\fnref{present-Carleton}}
\author[UHM]{J.\,G.~Learned}
\author[UHM]{S.~Matsuno}
\author[UHM]{M.~Sakai}
\author[KSU,IPMU]{G.\,A.~Horton-Smith}
\author[Stanford]{K.\,E.~Downum\fnref{present-MHS}}
\author[Stanford]{G.~Gratta}
\author[UTK,IPMU]{Y.~Efremenko}
\author[UTK]{O.~Perevozchikov\fnref{present-LSU}}
\author[TUNL,UNC]{H.\,J.~Karwowski}
\author[TUNL,NCCU]{D.\,M.~Markoff}
\author[TUNL,Duke,IPMU]{W.~Tornow}
\author[UWisc,IPMU]{K.\,M.~Heeger\fnref{present-Yale}}
\author[UW]{J.\,A.~Detwiler}
\author[IPMU,UW]{S.~Enomoto}

\author[Nikhef,IPMU]{M.\,P.~Decowski}
\address[UCB]{Physics Department, University of California, Berkeley, CA 94720, USA}
\address[LBNL]{Lawrence Berkeley National Laboratory, Berkeley, CA 94720, USA}
\address[IPMU]{Kavli Institute for the Physics and Mathematics of the Universe (WPI), University of Tokyo, Kashiwa, Chiba 277-8583, Japan}
\address[Tohoku]{Research Center for Neutrino Science, Tohoku University, Sendai 980-8578, Japan}
\address[OsakaGradSci]{Graduate School of Science, Osaka University, Toyonaka, Osaka 560-0043, Japan}
\address[UA]{Department of Physics and Astronomy, University of Alabama, Tuscaloosa, AL 35487, USA}
\address[Caltech]{W. K. Kellogg Radiation Laboratory, California Institute of Technology, Pasadena, CA 91125, USA}
\address[CSU]{Department of Physics, Colorado State University, Fort Collins, CO 80523, USA}
\address[Drexel]{Physics Department, Drexel University, Philadelphia, PA 19104, USA}
\address[UHM]{Department of Physics and Astronomy, University of Hawaii at Manoa, Honolulu, HI 96822, USA}
\address[KSU]{Department of Physics, Kansas State University, Manhattan, KS 66506, USA}
\address[Stanford]{Physics Department, Stanford University, Stanford, CA 94305, USA}
\address[UTK]{Department of Physics and Astronomy, University of Tennessee, Knoxville, TN 37996, USA}
\address[TUNL]{Triangle Universities Nuclear Laboratory, Durham, NC 27708, USA} 
\address[Duke]{Department of Physics, Duke University, Durham, NC 27708, USA}
\address[NCCU]{Department of Mathematics and Physics, North Carolina Central University, Durham, NC 27707, USA}
\address[UNC]{Department of Physics and Astronomy, University of North Carolina, Chapel Hill, NC 27599, USA}
\address[UWisc]{Department of Physics, University of Wisconsin, Madison, WI 53706, USA}
\address[UW]{Center for Experimental Nuclear Physics and Astrophysics, University of Washington, Seattle, WA 98195, USA}
\address[Nikhef]{Nikhef and the University of Amsterdam, Science Park, Amsterdam, Netherlands}
%
\fntext[present-KamObs]{Present address: 
Kamioka Observatory, Institute for Cosmic Ray Research, University of Tokyo, Hida, Gifu 506-1205, Japan}
\fntext[present-IPMU]{Kavli Institute for the Physics and Mathematics of the Universe (WPI), University of Tokyo, Kashiwa, Chiba 277-8583, Japan}
\fntext[present-OsakaGradSci]{Graduate School of Science, Osaka University, Toyonaka, Osaka 560-0043, Japan}
\fntext[present-UCD]{Physics Department, University of California, Davis, CA 95616, USA}
\fntext[present-LLNL]{Lawrence Livermore National Laboratory, Livermore, CA 94550, USA}
\fntext[present-UCLA]{Department of Physics and Astronomy, University of California, Los Angeles, CA 90095, USA}
\fntext[present-LBNL]{Lawrence Berkeley National Laboratory, Berkeley, CA 94720, USA}
\fntext[present-JLab]{Thomas Jefferson National Accelerator Facility, Newport News, VA 23606, USA}
\fntext[present-BNL]{Brookhaven National Laboratory, Yaphank, NY 11980, USA}
\fntext[present-UHM]{Department of Physics and Astronomy, University of Hawaii at Manoa, Honolulu, HI 96822, USA}
\fntext[present-Arcadia]{Department of Chemistry and Physics, Arcadia University, Glenside, PA 19038, USA}
\fntext[present-Carleton]{Department of Physics, Carleton University, Ottawa, Ontario K1S 5B6, Canada}
\fntext[present-MHS]{Milpitas High School, Milpitas, CA 95035, USA}
\fntext[present-LSU]{Department of Physics and Astronomy, Louisiana State University, Baton Rouge, LA 70803, USA}
\fntext[present-Yale]{Department of Physics, Yale University, New Haven, CT 06511, USA}

\begin{abstract}
We describe a compact, ultra-clean device used to deploy radioactive sources along the vertical axis 
of the KamLAND liquid-scintillator neutrino detector for purposes of calibration. The device worked by
paying out and reeling in precise lengths of a hanging, small-gauge wire rope~(cable); an assortment 
of interchangeable radioactive sources could be attached to a weight at the end of the cable. All 
components exposed to the radiopure liquid scintillator were made of chemically compatible 
UHV-cleaned materials, primarily stainless steel, in order to avoid contaminating or degrading the 
scintillator. To prevent radon intrusion, the apparatus was enclosed in a hermetically sealed housing 
inside a glove box, and both volumes were regularly flushed with purified nitrogen gas.  An infrared 
camera attached to the side of the housing permitted real-time visual monitoring of the cable's motion, 
and the system was controlled via a graphical user interface.
\end{abstract}


\end{frontmatter}

\newpage

\newpage
\noindent

\section{Introduction}
\label{sec:intro}

\begin{figure}[tp!]
\centering
\includegraphics[scale=0.52]{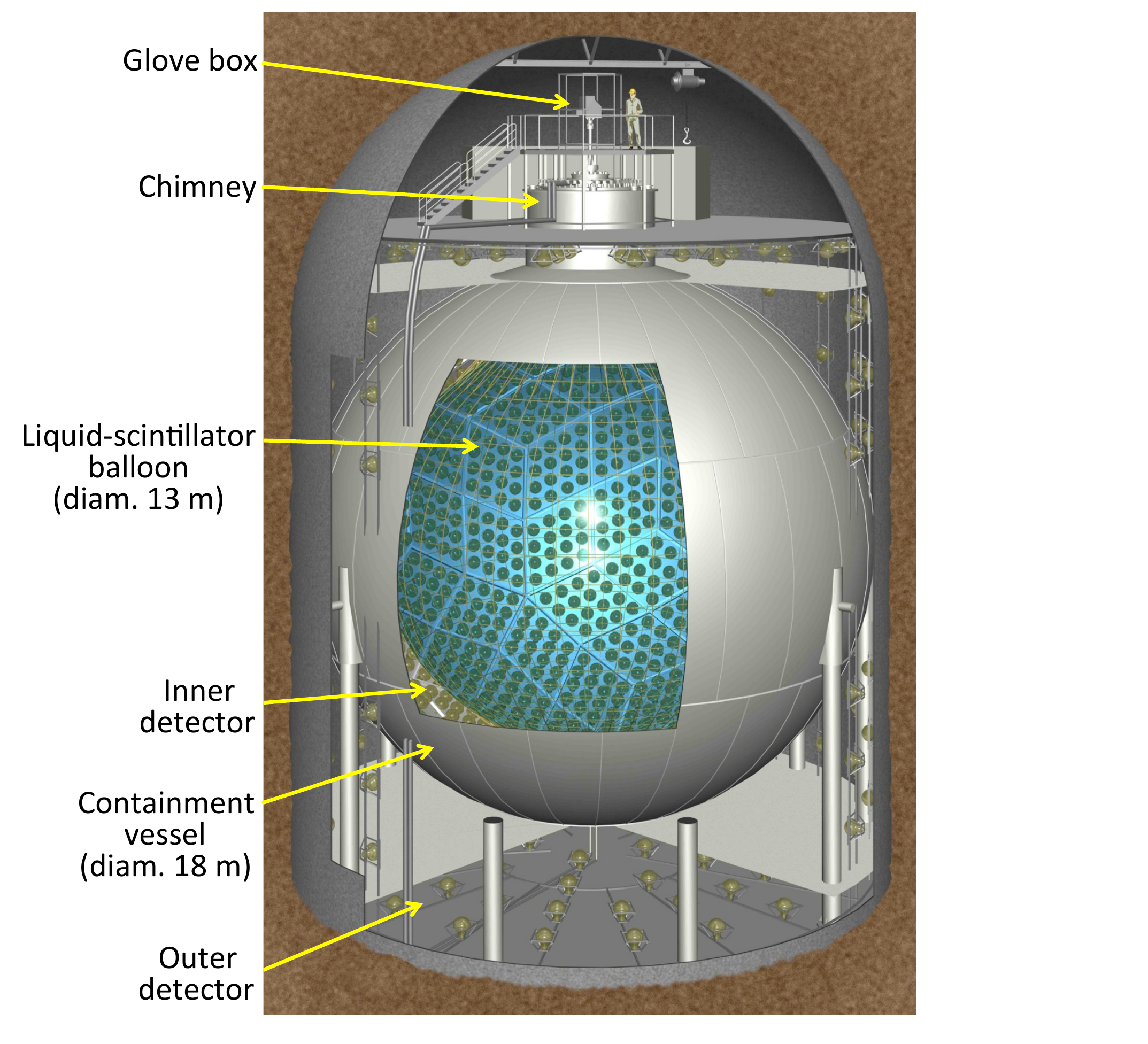}
\caption{The KamLAND detector.} 
\label{fig:kamland}
\end{figure}

The Kamioka Liquid-scintillator Anti-Neutrino Detector~(KamLAND) is a physics experiment
located 1~km underground in a mine near Kamioka-cho, Gifu, Japan.  
KamLAND~(Fig.~\ref{fig:kamland}) contains 1~kton of ultra-pure liquid scintillator~(LS) which 
serves as both the target and detecting medium for neutrino interactions.  A detailed 
description of the detector is given in~\cite{Abe:2009aa}.

KamLAND was commissioned in 2001 and used primarily to study the flux and energy spectrum 
of electron anti-neutrinos emitted by surrounding nuclear reactors.   The experiment provided 
convincing evidence that anti-neutrinos are massive particles that undergo flavor oscillations 
during propagation and it yielded precision determinations of the neutrino oscillation 
parameters~\cite{Eguchi:2002dm, Araki:2004mb, Abe:2008ee, Gando:2010aa}.  In addition, 
KamLAND was the first experiment to report evidence of geoneutrinos, which are anti-neutrinos 
produced by radioactivity in the Earth's interior~\cite{Araki:2005qa}. A subsequent measurement 
by KamLAND indicated that decay of radiogenic isotopes accounts for roughly half of the heat flux 
emanating from the Earth's interior; this was the first geoneutrino result to constrain geothermal 
models~\cite{Gando:1900zz}.

From May~2007 to February~2009 the KamLAND Collaboration conducted an intensive campaign 
to further purify the detector's LS, with the goal of enabling detection of solar neutrinos from 
$^7$Be decay in order to test aspects of the Standard Solar Model~\cite{Keefer:2013, 
KeeferPhDThesis, Gando:2014wjd}.
Purification was necessary because monoenergetic $^7$Be solar neutrinos are detected via 
elastic scattering by electrons, a method that requires much lower backgrounds than the 
delayed-coincidence technique used to detect higher-energy reactor anti-neutrinos.
In order to render a $^7$Be solar neutrino measurement possible, the already low concentrations 
in the LS of certain unwanted, long-lived isotopes---primarily $^{40}$K, $^{85}$Kr, and 
$^{210}$Pb---had to be reduced by several orders of magnitude.

Detector calibrations---in which radioactive sources of known energy are deployed 
to precise positions inside the LS in order to determine the timing of the surrounding 
photomultiplier tubes, the uncertainty in reconstructed event positions, and the 
detector's energy response---were identified as a potential threat to the radiopurity
improvements expected from LS purification.
There was strong evidence that the existing calibration systems had introduced 
contamination (primarily $^{222}$Rn, a highly mobile noble gas which generates 
the problematic long-lived daughter $^{210}$Pb) in amounts that would be 
unacceptable under the new experimental circumstances~\cite{Busenitz:2009ac}. 
Given that routine calibrations would be essential to the collection of high-quality 
data, the collaboration decided to implement a new, ultra-clean source deployment 
system for use following the LS purification campaign. The new calibration system, 
nicknamed ``MiniCal" due to its compact size, is the subject of this paper.

\section{System design}
\label{sec:design}

\subsection{Requirements and constraints}
\label{ssec:reqs-constraints}

The primary challenge in designing and building MiniCal lay in meeting 
oft-competing requirements regarding its size, cleanliness, and capabilities.

\begin{enumerate}
\item {\bf Compact size:} Every calibration system used in KamLAND has been installed 
inside a glove box atop the detector's chimney (Fig.~\ref{fig:kamland}).  Access to the
chimney and detector was possible via an 8-inch flanged opening in the floor of the glove box.
During normal data taking, detector access was blocked by two hermetically 
sealing gate valves located in series along the chimney's column, and the glove box was 
typically overpressured with nitrogen gas to prevent intrusion of contaminants, 
primarily radon.

Prior to MiniCal, hardware from a previous, full-detector-volume calibration 
system~\cite{Busenitz:2009ac} occupied a large fraction of the space in the glove box.  
We wanted to leave that system intact to preserve the possibility of making future off-axis 
deployments, so MiniCal was designed to fit in the small space available beneath 
it~(Fig.~\ref{fig:CAD-glovebox}).  
\begin{figure}[tp!]
\centering
\includegraphics[scale=0.60]{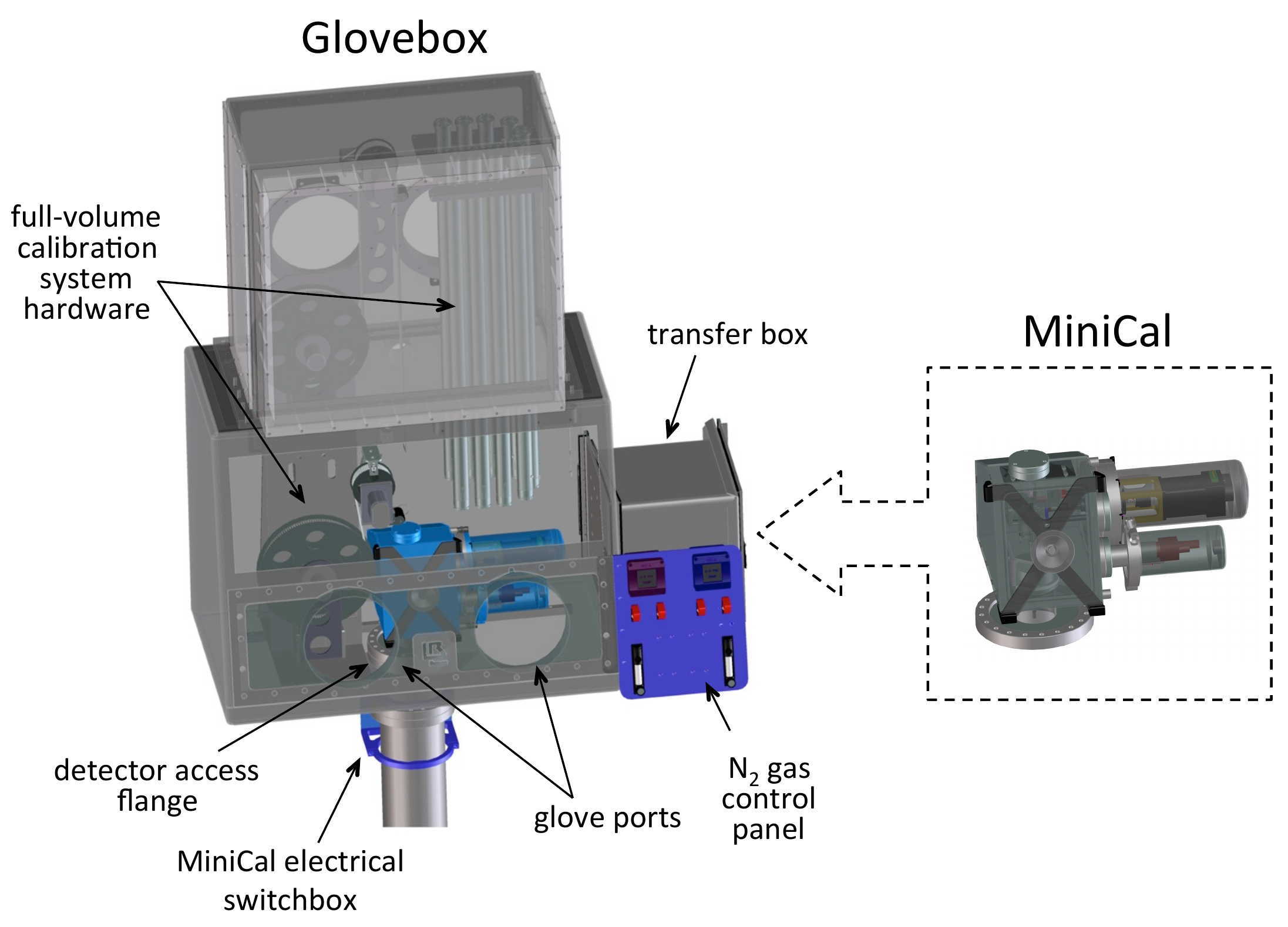}
\caption{Drawings of the MiniCal calibration device alone~(right) and installed inside 
the glove box atop the KamLAND detector~(left). MiniCal is highlighted in blue 
in the latter for visibility. Space inside the glove box was limited due to the continued 
presence of hardware from the previous, full-detector-volume calibration system.}
\label{fig:CAD-glovebox}
\end{figure}
It should be mentioned that the full-volume calibration system could not be deployed 
while MiniCal was installed because the latter blocked access to the detector.

The largest components of MiniCal also needed to be small enough to fit inside the glove 
box's 1-cubic-foot transfer box so that the new system could be moved into the glove box 
piecemeal and assembled inside.
And despite its small size, MiniCal needed to be amenable to servicing via the glove box's 
thick Viton gloves.
\vspace{\baselineskip}

\item {\bf Materials and cleanliness:} The KamLAND LS is a highly active, radiopure 
solvent composed primarily of dodecane and pseudocumene~\cite{Abe:2009aa}. All 
MiniCal components to be submerged in the LS had to be ultra-clean and possess low 
intrinsic radioactivity (so as to avoid contaminating the detector), and they had to be 
made of materials that would not be damaged by the LS or degrade its optical purity.
The MiniCal hardware in the glove box needed to be capable of withstanding prolonged 
exposure to LS vapors without suffering negative effects or emitting contaminants.

The two preceding calibration systems had used large-gauge wire rope and wide-ribbon 
woven-nylon cables, respectively, and the large surface areas of these materials tended 
to collect radon and radon daughters during idle periods and then transport them into the 
detector during deployments.  To reduce this problem---and also to meet the stringent 
spatial constraints---it was necessary that MiniCal utilize a small-gauge cable of some 
kind. We also elected to enclose the apparatus within its own hermetically sealed 
housing to provide a secondary barrier against radon intrusion.
\vspace{\baselineskip}

\item {\bf Accuracy and precision:} The KamLAND vertex reconstruction algorithm is 
tuned to minimize the difference between the nominal position of a deployed source 
and the mean of its corresponding event distribution, for a number of different source 
positions. Consequently, a systematic error in the positioning of sources---even if small 
compared to the detector resolution of $\sim$~12~cm/$\sqrt{E \rm (MeV)}$---can introduce 
a vertex reconstruction bias that significantly increases the systematic uncertainty in the 
fiducial volume estimate and thus negatively impacts the final physics 
result~\cite{Busenitz:2009ac}. Prior to MiniCal, the systematic error due to vertex 
reconstruction bias along the detector's central vertical axis (z-axis) was $\sim$~2~cm. 
We aimed to strongly limit the new deployment system's contribution to this error by 
making it capable of placing a source within $\pm2$~mm of any point along the z-axis, 
a length spanning 19~m from the glove box to the bottom of the LS balloon.
\vspace{\baselineskip}

\end{enumerate}

\subsection{Materials selection and radioassay}

The KamLAND LS is a mixture of 20\% pseudocumene, 80\% dodecane, and 1.36~g/l 
of the fluor PPO.  Pseudocumene is a highly active organic solvent which can damage
the structural integrity of many materials, so care had to be taken when choosing 
components for the calibration system.

The MiniCal components to be submerged in the LS or exposed to its vapors were made 
almost exclusively from materials that had been identified in past tests by the 
KamLAND collaboration~\cite{Busenitz:2009ac} and others~\cite{Freedman:1984pa} as
being chemically compatible with the LS---namely, 304 and 316~stainless steel, gold, 
titanium, nylon, Teflon, Viton, and quartz. Only 416~stainless steel, whose magnetic 
properties were needed for one small component which would be submerged (see 
Sec.~\ref{ssec:safeguards}), required new examination. LS soak and light attenuation 
tests on 416 samples indicated the component would not negatively affect the LS, or 
vice versa.  All other MiniCal components to be submerged in the LS were made of 304 
or 316~stainless steel.

To guard against introducing radioactive contaminants into the KamLAND detector, 
strict protocols were followed in cleaning and certifying the parts of the calibration system
that would enter the detector volume or be exposed to LS vapors.  All such parts were 
UHV cleaned according to a standard protocol employed at Lawrence Berkeley National 
Laboratory~(LBNL)~\cite{LBNL-ENG-10156A} and the system was preassembled and 
packaged in a class 10,000 clean room at LBNL before shipment to the experimental site. 
Items that were to be submerged in the LS, or that would come into direct contact with 
components to be submerged in the LS, underwent an additional cleaning and certification 
procedure~\cite{Busenitz:2009ac} consisting of a series of heated ultrasonic cleanings in 
a weak solution of aqueous trace-metal-grade nitric acid and deionized water (0.2~mol/l).  
A 100-ml sample of the nitric solution was collected at the end of the final cleaning and 
counted using a surface Ge detector operated by the LBNL Low Background 
Facility~\cite{LBF}. No evidence of U, Th, or K contamination above the detector's sensitivity 
limits was observed; the upper limits after a $\sim$~30-day count were U~$<4$~mBq/kg, 
Th~$<2$~mBq/kg, and K~$<130$~mBq/kg, where kg refers to leachate mass.

\subsection{Hardware}
\label{ssec:hardware}

\begin{figure}[t!]
\centering
\includegraphics[scale=0.54]{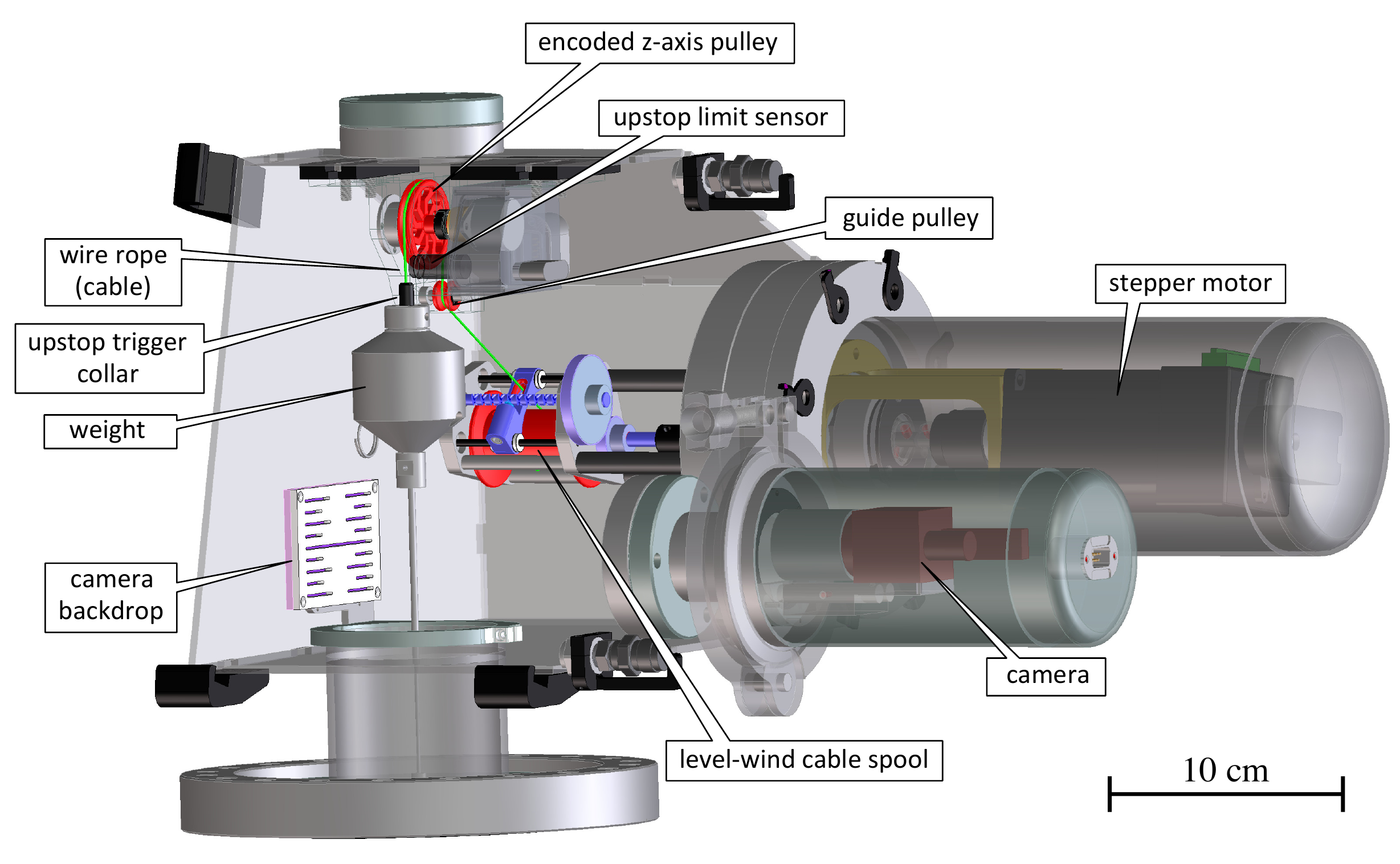}
\caption{Cutaway drawing of the interior of the MiniCal housing. The cable is drawn in
green and the components it touches are shaded red. The driveshaft and level-wind 
mechanism are shaded blue.}
\label{fig:minical-interior-photo}
\end{figure}

The basic concept for the MiniCal system is simple: it used a stepper motor to turn a spool 
of cable and thereby lower and raise a radioactive source along the z-axis of the KamLAND 
detector (Figure~\ref{fig:minical-interior-photo}).
The motion of the cable was measured by an encoded pulley located above the z-axis, 
enabling precise positioning of the source.

We elected to use a thin-gauge cable because compared to the alternatives (e.g., a chain 
or sprocketed cable) it offered the best combination of simplicity and cleanliness.  The 
primary drawback was that great care had to be taken to prevent cable slippage on the 
encoded pulley, as that would have ruined positioning accuracy.
The cable was made by Strand Products, Inc., from \nicefrac{1}{32}-inch-diameter, 
7$\times$19-stranded, 120-pound tensile wire rope that had been proof-loaded to 60\% 
of its breaking strength to eliminate constructional stretch\footnote{Our own dynamometer 
tests demonstrated that the breaking strength of the cable assemblies was approximately 
100~pounds (45~kg).}.  This particular cable was selected for several reasons: it is 
composed entirely of T304 stainless steel, which is LS-compatible; it has a small diameter, 
which served to minimize both the amount of space it occupied when spooled and the
amount of surface area that could collect contaminants; and its $7\times19$ strand lay renders 
the cable very supple and allowed it to maintain good contact with the encoded z-axis pulley, 
which was machined from titanium to minimize its rotational inertia.
Each end of the cable terminated in a swaged bushing: one end was anchored to the spool,
while the other end was attached to a free-hanging 0.55~kg weight which kept the cable 
under tension to ensure good contact with the encoded pulley.  A captured, spring-loaded 
pin in the underside of the weight was used for easy attachment of different radioactive sources.
The attachment mechanism was designed to be backwards-compatible with existing sources 
used with the previous calibration systems, requiring only that each source be fitted with an
adaptor bolt.

The cable was wound on a custom spool modeled after a Penn GTO~220 level-wind 
fishing reel.  This device contained a cable guide mechanically coupled to the spool's rotation 
to wind cable evenly and compactly along the length of the spool.
A small guide pulley located between the spool and the z-axis pulley prevented the 
side-to-side motion of the spooling cable from affecting its smooth passage over the z-axis 
pulley.  The spool was turned by an Anaheim Automation 23MDSI integrated stepper motor 
performing 1600~steps/revolution.

\begin{figure}[tp!]
\centering
\includegraphics[scale=0.60]{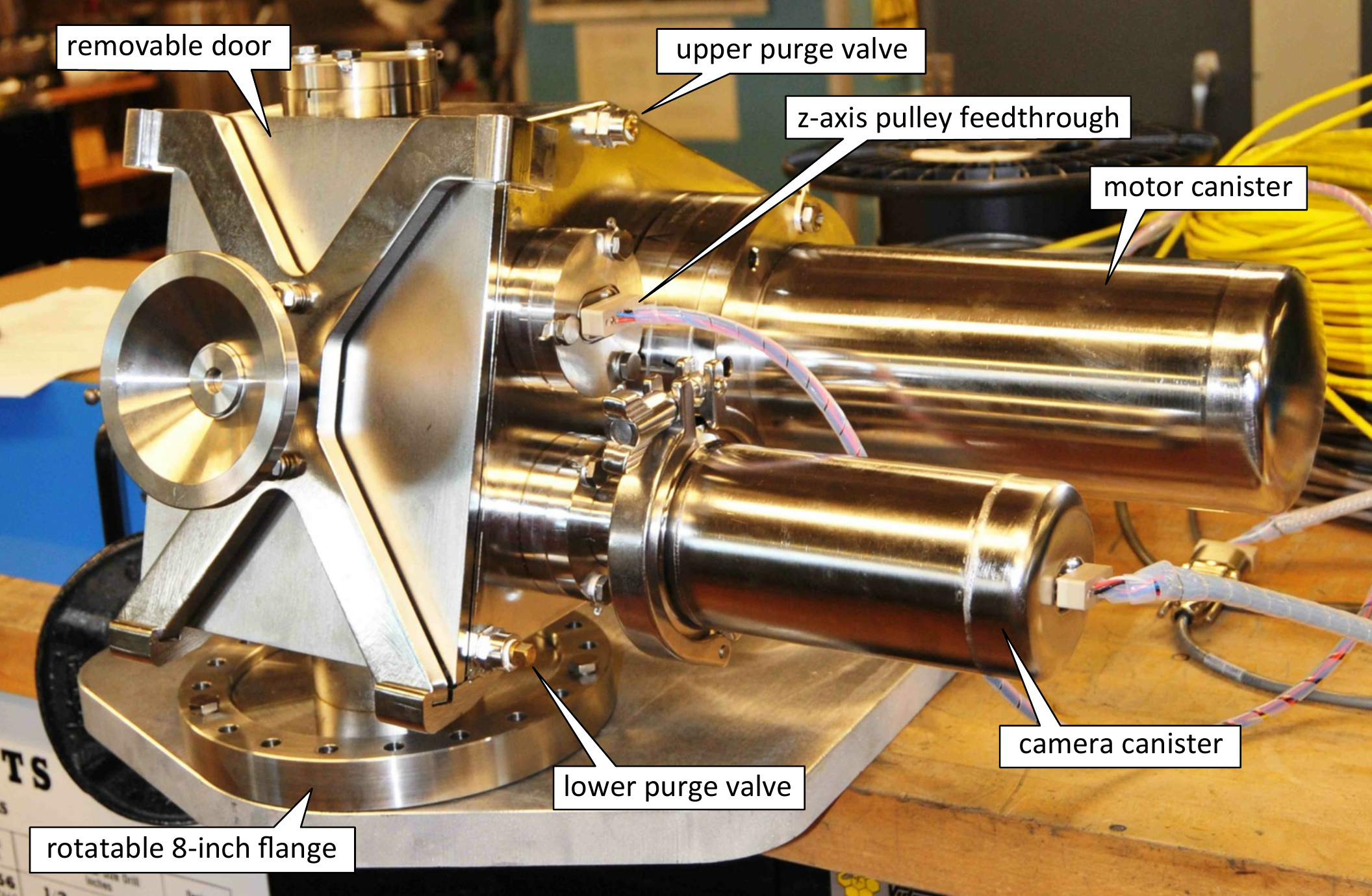}
\caption{The MiniCal housing in a test bench setup.}
\label{fig:minical-test-bench-photo}
\end{figure}

It had been observed that the previous KamLAND calibration system had introduced 
small amounts of radon into the LS, likely due to small leaks in the glove box that allowed
ingress of ambient outside air.  In order to provide an added barrier against radon intrusion, 
the MiniCal hardware was enclosed within a hermetically sealed housing 
(Figure~\ref{fig:minical-test-bench-photo}) which was bolted to the 8-inch-diameter conflat 
flange welded to the floor of the KamLAND glove box (Figure~\ref{fig:CAD-glovebox}).  
The housing was constructed by welding together jointed, $\nicefrac{1}{8}$-in.-thick 
stainless steel panels; afterwards it was electropolished to remove oxidation and 
passivate the surface.  The interior was accessed through a 5.5$\times$7.4~in.$^2$ 
rectangular opening which faced the glove box's gloves and was usually covered by a 
removable door.  A Viton O-ring seated in the face of the door formed a seal against the 
housing when the the door's hand-operated dial was cranked to press the two surfaces 
together.  The housing sat atop a 1.8-in.-tall stand tube (3-in.~OD, 2.75-in.~ID) 
in order to provide clearance for inserting bolts into the 8-in.\ flange at its base.

The z-axis pulley was mounted to the housing's ceiling on a movable stage which could 
be translated in two dimensions to center the hanging cable over the detector's vertical 
axis. The centering procedure was performed at the time of installation with assistance 
from a cross-hair-engraved cover plate inserted in the stand-tube opening inside the
MiniCal housing.

The floor of the housing was angled slightly so that LS brought up by the cable 
during deployments collected in its front.  The accumulated LS---which was prevented 
from returning to the detector by a small collar around the stand-tube opening in the 
housing's floor---was periodically drained through a manually operated Swagelok Whitey 
P-series stainless steel purge valve (with PTFE ball) mounted on the lower corner of the
housing.  This valve and an identical one located above it also functioned as vents 
during nitrogen gas purges of the MiniCal housing volume (see Sec.~\ref{ssec:src-exchanges}).  
The spring-loaded valves were bracketed by steel retaining arms to prevent them from 
being unscrewed to the point of disassembly.

In order to prevent the stepper motor from introducing contaminants or coming into 
contact with corrosive LS vapors, it was enclosed inside a leak-tight stainless steel 
canister attached to the side of the housing.  The motor's 1/4-in.\ 
driveshaft connected to the cable spool via an MDC Vacuum direct-drive rotary 
mechanical feedthrough.

An infrared camera provided real-time visual monitoring of the cable when the housing
was sealed shut during deployment operations.  Like the motor, the camera---an Allied 
Vision Technologies Guppy F-038B black-and-white, near-infrared, asynchronous 
Firewire digital camera with a Kowa Model LM25JC 2/3-in.\ machine vision lens---was 
enclosed in a sealed canister.  The camera viewed the interior of the housing through 
a quartz window, with two Advanced Illumination SL4301 880-nm IR LED 5-volt spot 
lights attached to the underside of the camera providing illumination.  Infrared imaging
equipment was chosen because its operating spectrum lies above the 
$\sim$~600~nm optical cutoff of the detector's PMTs.

All MiniCal flanges were made of stainless steel and joined using titanium bolts 
to avoid galling.  The bolts were tethered to prevent being dropped or lost during 
assembly and disassembly.  The flanges did not rely on knife edges to form gas-tight 
seals; instead, they had been modified to employ seated Viton O-rings, based on a 
design developed at LBNL.  This technique offers several advantages over conventional 
flange gasket seals, which are single-use and can be awkward to manipulate manually.  
A rotatable 8-in.\ flange was used for the base of the MiniCal housing to permit 
convenient alignment of the housing during installation. The camera canister contained 
the assembly's lone quick flange, which was useful for providing prompt access to the 
camera during its initial tuning.

\subsection{Motion control software}

\begin{figure}[t!]
\centering
\includegraphics[scale=0.45]{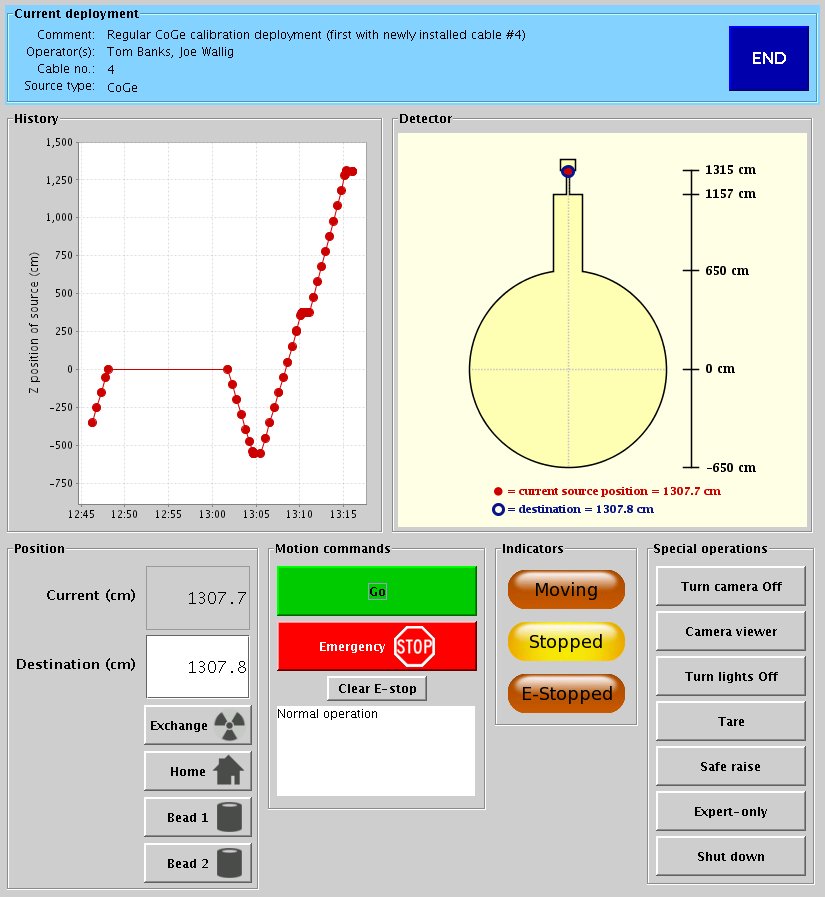}
\caption{MiniCal graphical user interface~(GUI).}
\label{fig:GUI-labeled}
\end{figure}

The MiniCal system was controlled and monitored using custom Java software running 
on a dedicated computer in an electronics room adjacent to the experiment.  The 
software consisted of two independently functioning parts: a control side which 
communicated with the hardware, and a graphical user interface, or GUI 
(Figure~\ref{fig:GUI-labeled}), which communicated with the control side.  
The two layers used the Java Remote Method Invocation protocol to communicate 
with each other in real time via changes to the values of a set of shared parameters.  
All system actions and parameter changes were logged to a PostgreSQL database.

The computer, a small-form-factor CappuccinoPC SlimPRO 620S, ran the open-source 
Fedora~9 Linux operating system.  Three 9-pin serial ports were used to communicate with 
the stepper motor and encoders, and a Firewire port was used to communicate with the 
camera.  The camera's output was viewed using the software program Coriander.  
One of the serial ports was used to communicate with both the stepper motor and a Crouzet 
XT20 Millenium II+ programmable logic controller~(PLC) which enabled software-controlled 
switching of power to the hardware.  The PLC was used primarily to power the hardware 
on and off at system startup and shutdown, and to toggle power to the driveshaft safety 
brake and the IR lights during system operation.

The stepper motor's rotation was measured by a US Digital E5D optical encoder which 
was read out by a US Digital AD4-B-D quadrature encoder counter.  The motor's 
revolutions could not be used to precisely determine cable payout, however, because 
the spool's effective radius changed with the amount of spooled cable.  Instead, 
cable payout (and hence the source's position) was determined using the 3.6-cm-diameter 
titanium z-axis pulley, whose rotations were measured by a Renishaw RM22I 
non-contact rotary encoder with a resolution of 800~counts/revolution.  The z-axis pulley 
was attached to a gold-plated rare-earth magnetic actuator which rotated inside the 
fixed stainless steel RM22I encoder housing; the encoder transmitted standard 
incremental quadrature output which was counted by another US Digital AD4-B-D unit.
The pulley encoder electronics were enclosed inside a small hermetically sealed stainless 
steel box inside the MiniCal housing.

Because cable payout was controlled by the motor but measured by the z-axis pulley, 
the system generally had to iterate several motions before the deployed source 
reached the desired destination.  When the user instructed the system to move the source 
to a new destination, the software computed a conservatively low estimate for the number 
of motor steps required and programmed this number into the motor. After the stepper 
motor executed the programmed index, the software compared
the source's current position against the destination and reprogrammed the motor as 
necessary.  This process was repeated until the source converged to within 0.1~mm 
of its destination.  The maximum translational speed of the cable during motion
was restricted to $\sim$~3.5 cm/s.  The software took into account the fact that the
amount of cable stretch varied slightly according to the weight of the attached 
source and the length of cable paid out\footnote{We characterized the cable's stretching 
behavior by measuring lengths of a similar cable when hung with loads of 0~kg, 0.5~kg, 
and 1.0~kg.}. Most of the sources generated a maximum additional stretch of only 0.15~cm, 
though for one unusually heavy source the value was 0.6~cm.

The computer and power supplies were physically separated from the hardware by 
roughly 130~ft of distance and several barriers, chiefly the glove box and the 
MiniCal housing.  To span the gap, the system utilized two intervening custom-built 
electronics boxes, one at each end, which contained devices for transmitting and 
processing data and power.  The stepper motor and the two encoders communicated 
in RS485 protocol, which is well-suited for long-distance transmission; the motor and
encoder signals were converted to RS232 protocol by a special adaptor and encoder 
counters, respectively, to communicate with the computer's serial ports.  The 
camera's communications were transmitted using a Newnex FireNEX-CAT5 S400 
extender pair, which converted the short-range IEEE~1394a (Firewire) protocol into a 
form suitable for long-distance transmission over standard cat5e ethernet cable at 
400~Mbps.  CeramTec sub-D and micro-D weldable hermetic feedthroughs were used 
to pass the electrical signals and power through the walls of the glove box and the 
MiniCal housing.  Radiopurity considerations prohibit the use of solder in components 
exposed to the glove box environment, so all in-glove box electrical wiring was made 
from Teflon-coated wires and PEEK connectors using gold-plated crimped connections.

\subsection{Safeguards}
\label{ssec:safeguards}

The MiniCal system possessed several built-in safety features, both in hardware
and software, to prevent accidents or damage to either the detector or to the system 
itself.
\begin{itemize}
\item An Anaheim Automation NEMA23 electromagnetic friction brake was attached 
	to the driveshaft between the stepper motor and the spool.  When electrical 
	power was applied to the brake its electromagnet overcame the mechanical
	force from a spring, permitting the driveshaft to turn freely; when power was turned
	off the spring forced friction plates together, clamping the driveshaft in place.  
	This type of ``power-off" or ``fail-safe" brake was desirable because it would
	immediately lock the driveshaft if power were lost.
\item A Polyclutch EFS 16 ``slip-ease'' clutch was connected to the driveshaft between 
      the motor and the brake.  In the unlikely event that the cable became caught, 
      the clutch slipped at 4~kg weight equivalent, far below the cable's approximately 
      45~kg breaking strength.  Moreover, the motor itself slipped at roughly 4.5~kg 
      weight equivalent, so the system was incapable of generating sufficient tension 
      to break the cable.
\item An Automation Direct 12-mm round stainless steel inductive proximity 
	sensor was installed just below the z-axis pulley to prevent the weight from 
	being raised too high and colliding with the pulley superstructure.  The upstop 
	sensor was triggered by a small 416 stainless steel collar installed at the end 
	of the cable just above the weight.  When the sensor was triggered it sent a 
	signal directly to the motor that blocked its ability to execute further upward 
	motion.  (The motor remained capable of downward motion.)  
\item The cable length was chosen so that even at maximum extension an 
	attached source could not come into contact with the bottom of the KamLAND
	balloon.
\item The GUI enforced restrictions on acceptable user-entered values to prevent 
	the system from operating outside safe limits.
\item The software continuously monitored the status of the system, most notably 
	the motion of the motor and z-axis pulley.  If the software observed any motion 
	when the system was supposed to be stopped, or if the driveshaft or pulley rotation 
	speeds exceeded certain limits, or if there was any disagreement between the motor 
	encoder and the z-axis pulley encoder (e.g., due to cable slippage), the software 
	immediately stopped the motor and engaged the safety brake to prevent the cable 
	spool from turning.  The system would therefore self-arrest if the motor failed.
\end{itemize}
In the unlikely event that the motor failed during a source deployment, the source
could be retrieved by removing the motor assembly and replacing it with a special
ratcheted hand crank.  The software possessed a special emergency mode for 
this situation.

\section{Operating protocols}
\label{sec:operations}

The MiniCal system was designed to be safe, simple, and capable of operation in
all aspects by a single person.  Operators were responsible for exchanging sources, 
managing the flow of nitrogen gas through the system, and conducting calibration 
deployments.

\subsection{Source exchanges}
\label{ssec:src-exchanges}

The first step in preparing for a calibration was to attach the desired radioactive source 
to the end of the MiniCal cable.  
A list of the sources used and their energies is given in Table~\ref{tab:sources}.
\begin{table}[tp]
\centering
\begin{tabular}{cccc}
\hline \hline
\T \B Source & Half-life & Radiation & Energy~(MeV) \\
\hline
\T $^{203}$Hg & 46.6~d & $\gamma$ & 0.279 \\ 
    $^{137}$Cs & 30.1~y & $\gamma$ & 0.662 \\
    $^{68}$Ge & 271.0~d & $\gamma$ & 1.022 \\
    $^{65}$Zn & 243.9~d & $\gamma$ & 1.116 \\
    $^{241}$Am-$^{9}$Be & 432.6~y & n & \ 2.223$^\dagger$ \\
\B $^{60}$Co & 5.3~y & $\gamma$ & \ 2.506$^\ddagger$ \\
\hline \hline
\end{tabular}
{\normalsize \\
\T $^\dagger$\,From delayed capture by hydrogen. Some prompt $\gamma$ rays are also seen. \\
\B $^\ddagger$\,From successive 1.173~MeV and 1.333~MeV $\gamma$ rays. \hspace{1.33in}}
\caption{Radioactive sources used in detector calibrations, ordered by energy.  These 
particular sources were chosen because their radiation spans the low-energy range of the 
neutrino spectrum observed in KamLAND.}
\label{tab:sources}
\end{table}

\begin{figure}[tp!]
\centering
\includegraphics[scale=0.50]{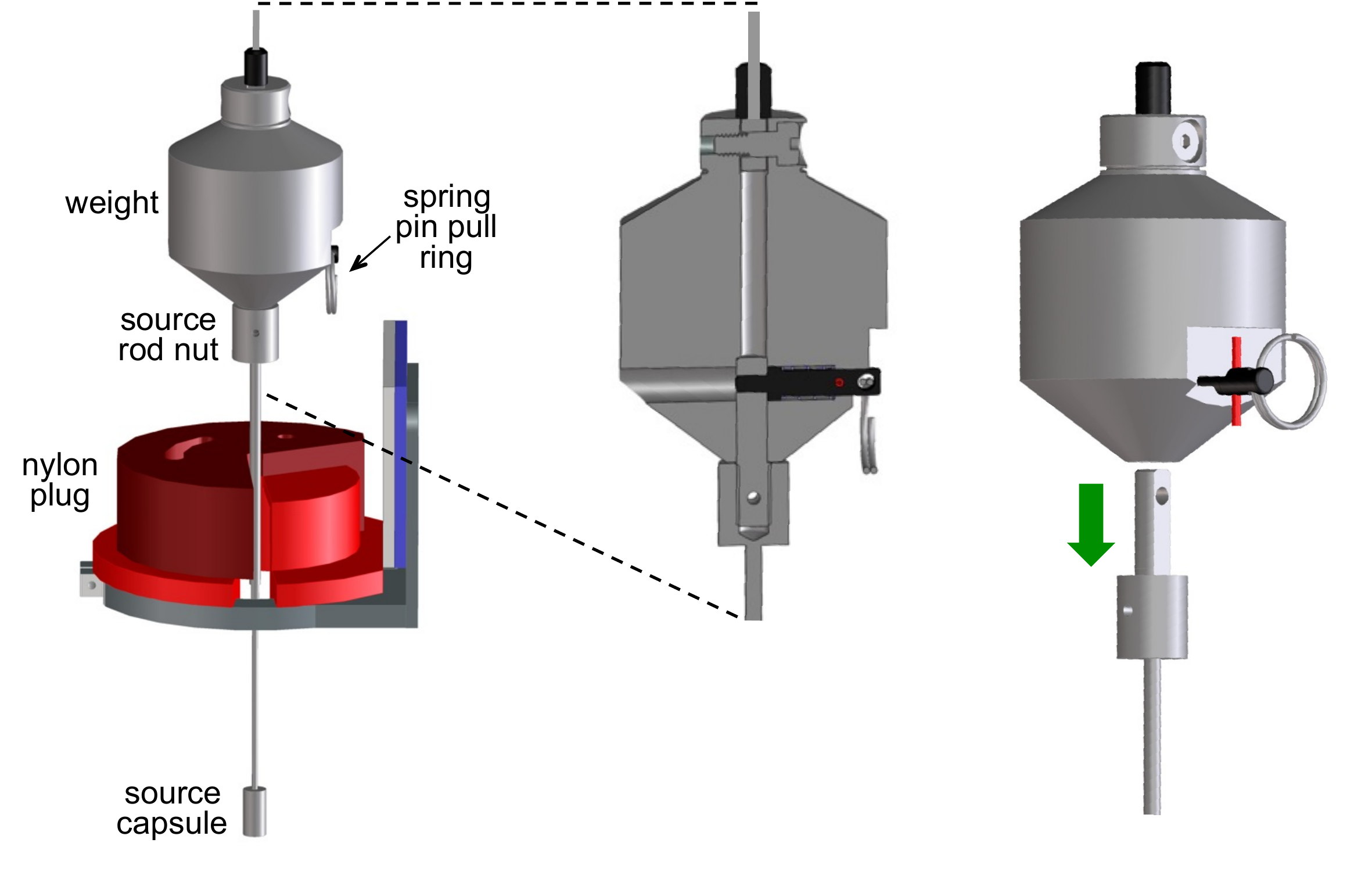}
\caption{Illustrations of the weight and its source attachment mechanism. An 
internal spring-loaded pin is used to capture source rods inserted into the weight; the 
source rods have a D-shaped profile that ensures correct alignment. The spring pin 
has two small tines (highlighted above in red) to provide an unambiguous visual 
indicator of when a source has been securely captured inside the weight, and for 
anchoring the spring-loaded pin in its retracted position. A nylon plug with a rotatable 
disk (shown above, in red, in its open position) was used for safely capturing a source 
released from the weight.}
\label{fig:src-weight-plug}
\end{figure}

To exchange sources, the weight and its attached source were first raised to the 
uppermost possible position inside the MiniCal housing, and the chimney's upper 
gate valve was closed as a safety measure.  The operator then removed the front 
access door and installed a special nylon plug in the floor opening of the housing.  
This plug contained a rotatable disk that could be closed around the source rod to 
securely capture it and prevent it from being dropped down the detector access tube.  
At this point the operator released the source from the weight by pulling the ring 
on the spring-loaded attachment pin, allowing the source to fall out and be caught 
by the plug (Figure~\ref{fig:src-weight-plug}).

To install a source, the procedure was reversed.  The source was lifted up by hand 
and inserted in the bottom of the weight; the source shaft's D-shaped cross section 
ensured alignment with the attachment pin.  Once the source was fully inserted, the 
spring-loaded attachment pin was released to capture the source inside the weight.  
The pin had two small tines that provided unambiguous visual confirmation of a 
successful attachment; when the tines disappeared from view inside the weight, 
the source and weight were securely connected and the plug could be removed.

After the source exchange was complete, the operator reinstalled the door of the
housing, reopened the chimney's upper gate valve, and configured the system so that 
purified nitrogen gas continually flowed up the chimney through the MiniCal housing 
and out of the glove box, to purge the volumes of any radon.  After several hours
had passed (typically by the morning of the following day), the operator ended the 
nitrogen gas purge, sealed the MiniCal housing off from the glove box by closing the 
purge valves, opened the chimney's lowermost gate valve to permit access to the 
detector, and proceeded with the source deployment.

\subsection{Calibration deployments}
\label{ssec:deployments}

Deployments were generally conducted ``remotely'' from the experiment's nearby control room, 
where the operator established a virtual network computing~(VNC) connection to the MiniCal 
computer to manipulate the system using the GUI and to monitor the live camera feed.

\begin{figure}[t!]
  \centering
  \subfloat[]{\label{fig:tare-image-photo}\includegraphics[scale=0.65]
  	{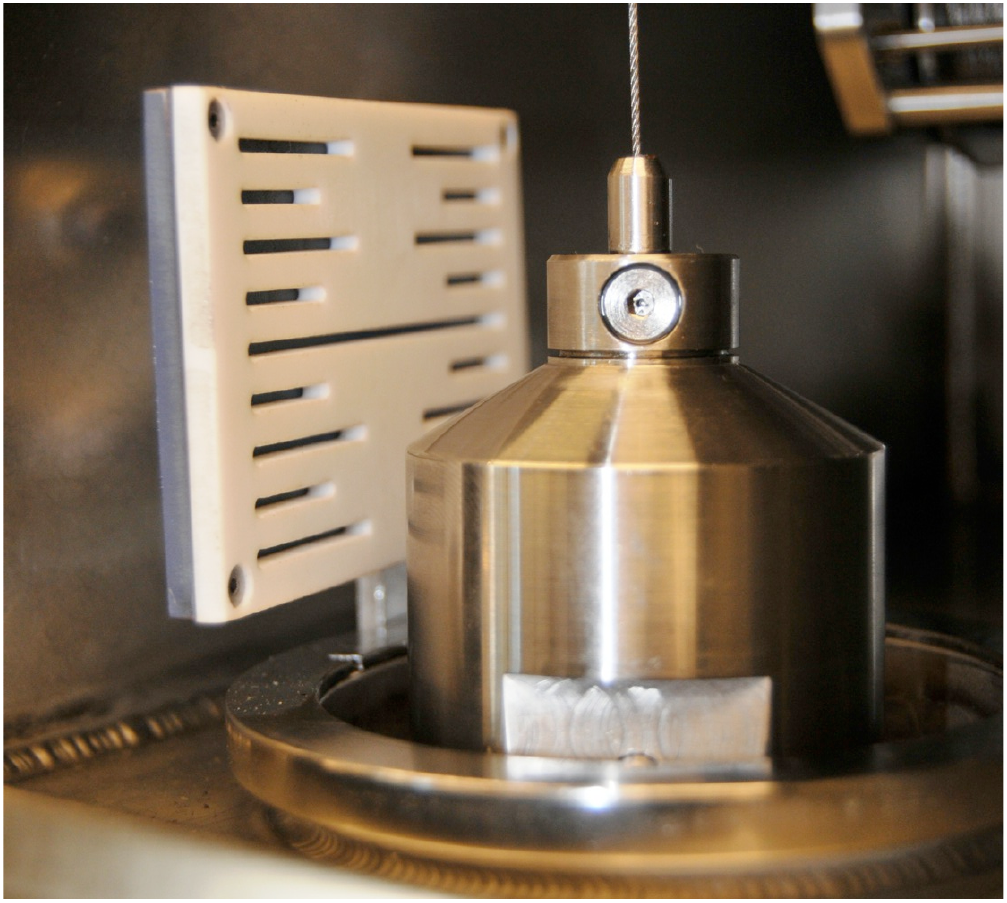}}
  \hspace{2mm}
  \subfloat[]{\label{fig:tare-image-camera}\includegraphics[scale=0.65]
  	{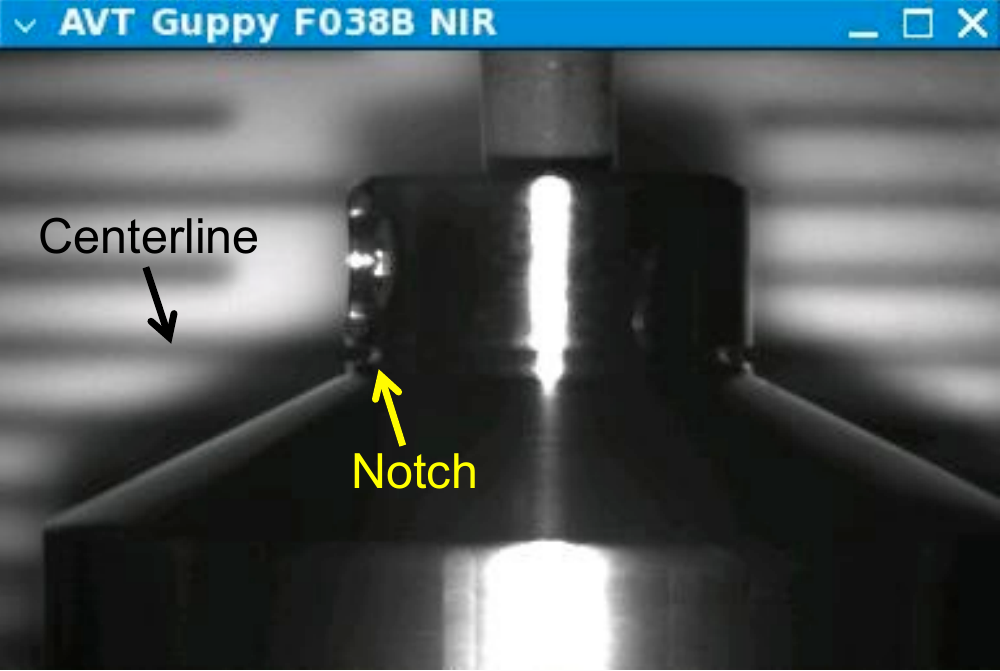}}
  \caption{Two images of the weight in its ``tare" position, in which the notch in the 
  weight's neck is aligned with the long centerline in the camera backdrop.
  (a) Photo of the weight and rulered backdrop inside the MiniCal housing.
  (b) The same scene as viewed through the camera.
  The backdrop's hash marks are spaced at 5~mm intervals.}
  \label{fig:tare-images}
\end{figure}

Before starting a deployment, the operator had to calibrate the system by zeroing the
source at a known position.  This was accomplished by moving the weight to the position
where a notch on its neck was visually aligned with a horizontal reference line in the center of 
the camera backdrop, a water-jetted sheet of Teflon attached to an electropolished titanium 
backing plate for contrast (Fig.~\ref{fig:tare-images}).
The operator then pressed the ``Tare" button in the GUI and the software calculated the 
source's center of activity, using the premeasured source lengths and the known location 
of the reference line with respect to the detector.

\begin{figure}[tp]
\centering
\includegraphics[scale=0.50]{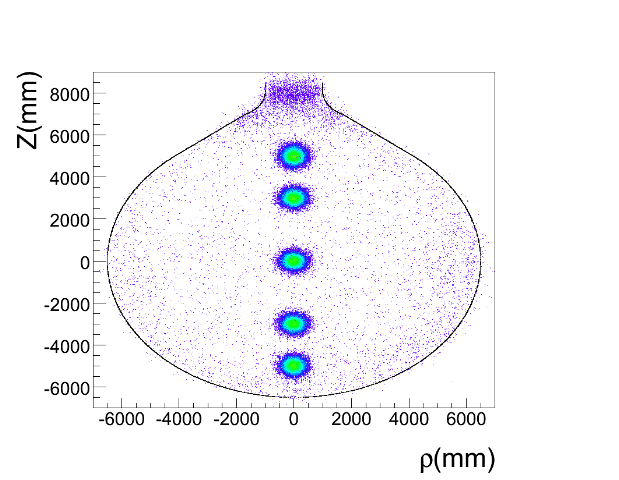}
\caption{Activity from a $^{60}$Co source deployed to 
${\rm z}=+5, +3, 0, -3, -5$~m along the detector's z-axis, as reconstructed from 
the data. The colors correspond to the intensity of detected activity.}
\label{fig:minical-deploy-data}
\end{figure}

After its position had been established by a tare operation, the source was typically
lowered to each desired position in succession, beginning with the uppermost position.  
When moving between positions, the operator watched the cable as it passed through 
the camera's field of view, in order to verify that the system was behaving correctly and 
to monitor the cable's condition (e.g., check for frayed wires).  
After the source came to a stop at a deployment position, the operator turned off the 
infrared lights, set the driveshaft brake for security, and commenced taking data.  
A separate data file was recorded for each calibration point 
(Figure~\ref{fig:minical-deploy-data}).

To enable consistency checks of the system's position, two small sleeves (``beads") 
were swaged onto the cable at the time of its manufacture, and the exact positions 
of the beads were found by hanging the cable under its expected load in a controlled 
environment at LBNL and measuring it with a precision tape measure. By comparing 
the known bead positions with the positions reported by the MiniCal system, the operator 
could verify the system was accurately reporting the cable payout; this was the only 
precise means of verifying the system's performance in situ.  The beads had a 
negligible effect on the system's measurement of cable payout.

Coarse assessments of the source's position and the detector's contamination levels
were provided in near real time by an automated online data analysis which processed
most of the incoming detector data and displayed the results.  The online software was 
configured for general-purpose analysis and the information it provided was useful 
primarily for monitoring for any gross abnormalities while a deployment was in progress.

Routine detector calibrations were typically performed once every two weeks.  
A complete source deployment lasted between 5--8 hours, depending upon the 
type and intensity of the source and the number of deployment positions.
For example, a standard calibration using a $\sim$~200~Bq $^{60}$Co
source consisted of placing it at 0.5-meter intervals between \mbox{${\rm z}=6.0$~m} 
and \mbox{${\rm z}=-6.0$~m} (where \mbox{${\rm z}=0$~m} is the center of the LS balloon) and leaving 
it at each point for $\sim$~15~minutes.

\section{Performance}
\label{sec:performance}

MiniCal was operated routinely from February~2009 until June~2011, when it was 
dismounted to permit a final deployment of the full-volume calibration system.
During its operational period MiniCal performed 65~source deployments and (including
tests) executed more than 90~round trips into the detector.
Immediately before being uninstalled, MiniCal was used to continuously deploy a CdWO$_4$ 
crystal assembly to the center of the detector for a special study lasting several days.

\subsection{Positioning accuracy and reproducibility}

Upon being installed in the KamLAND glove box, MiniCal exhibited near-perfect reproducibility
in positioning.  However, the cable payout ratio that had been established during system testing 
no longer appeared to be correct, as the position of the single bead reported by the system 
disagreed by 2.9~cm with prior tape measurements of the loaded hanging cable---that is, 
MiniCal reported the cable to be almost 3~cm shorter than we believed was the case. After 
accounting for the effects on cable stretch from buoyancy ($-0.1$~cm) and the precise value of 
the attached weight ($-0.7$~cm) there was still a persistent 2.1-cm discrepancy. 
It thus appeared the system's payout ratio had somehow been altered in the course of being 
disassembled, cleaned, reassembled, and installed in the glove box. We made the decision to 
recalibrate the payout ratio to the premeasured position of the bead; this entailed increasing
the payout ratio from 11.298~cm/rev to 11.315~cm/rev, which corresponds to an increase of 
0.0027~cm in the effective radius of the z-axis pulley.

In January 2011 the original single-beaded cable was replaced in situ with a similar cable
possessing two evenly spaced beads which provided an additional reference point over a longer 
length.  The positions of the two beads had been measured with care beforehand using a 
precision tape on the hanging cable under load.  The new, double-beaded cable verified the 
correctness of the recalibrated cable payout ratio established roughly two years earlier.  The 
reason for the change in the payout ratio between the system's offsite testing and onsite 
operation remains unknown.

A small hysteresis of 2--3~mm of undershoot was consistently observed when the weight 
returned to its tare position after being deployed to its maximum depth in the detector. This
was likely due to a small degree of cable slippage on the z-axis pulley from the lubricating 
effects of the LS during reel-in.  The effect was small, but to minimize the risk of positioning 
errors we typically placed a source at progressively lower positions during a calibration and 
only raised it at the conclusion of the deployment.

It was observed that small white crystals sometimes formed on the portion of the cable near 
the weight when the system was idled for weeks or more.  We believe the crystals were PPO 
from the LS brought up by the system, and their formation was likely promoted by the drying 
effects of the constant nitrogen gas flux through the MiniCal housing.  We observed that the 
crystals dissolved immediately upon submersion in the LS and did not affect the cable payout 
accuracy.

\subsection{Contamination assessment}

The KamLAND detector itself provides the most sensitive means of measuring radioactivity
inside it. The level of $^{222}$Rn is of particular interest because it was known to have been 
introduced into the detector by previous calibration systems, and its long-lived daughter 
$^{210}$Pb would have presented a problematic background to detecting $^7$Be solar 
neutrinos during the high-purity phase of the experiment.

\begin{figure}[tp!]
\centering
\includegraphics[scale=0.75]{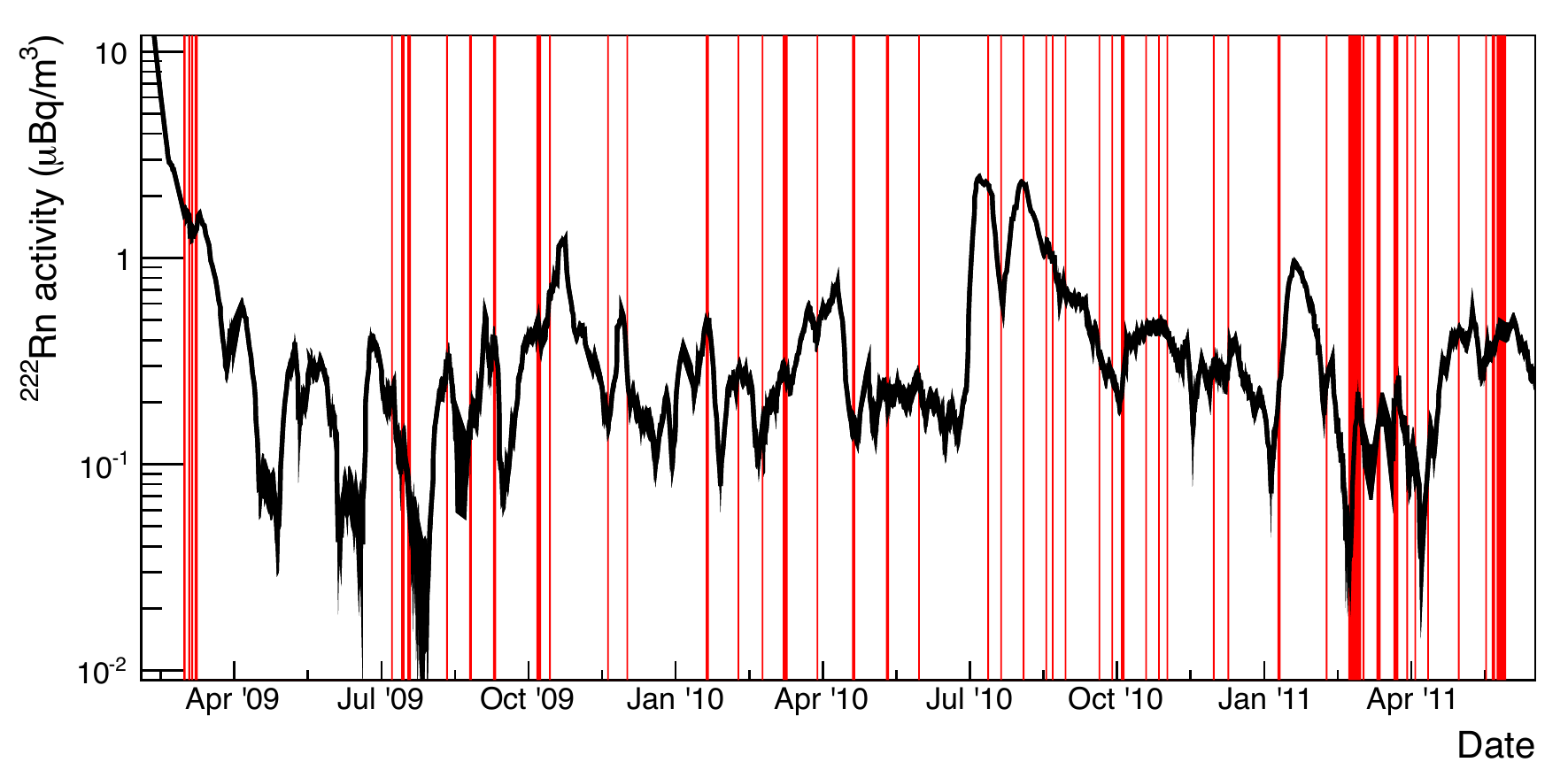}
\caption{$^{222}$Rn activity in the KamLAND detector over the period MiniCal was in use.
The radon activity level is determined from the efficiency-corrected number of $^{214}$Bi--$^{214}$Po 
decay coincidences inside a 5.5-m-radius spherical volume; the statistical error is 
reflected in the width of the black line.  The vertical red bands indicate MiniCal deployments 
into the detector; no correlation between radon activity and deployments is apparent.  
The prominent feature beginning in July~2010 is due to a disruption in the 
detector's water cooling system which produced strong convection in the LS and brought 
contaminants from the balloon's surface into the fiducial volume.}
\label{fig:radon-vs-time}
\end{figure}

The level of $^{222}$Rn activity inside the LS can be determined via its daughters
$^{214}$Bi and $^{214}$Po, whose successive decays occur close together in time 
($\sim$~1~ms) and therefore generate a robust delayed-coincidence signal. 
The measured $^{222}$Rn activity level vs.\ time is plotted in Fig.~\ref{fig:radon-vs-time}
for the period when MiniCal was in use. 
As the figure shows, there is no apparent correlation between MiniCal deployments and 
changes in $^{222}$Rn activity. Indeed, most of the features in the $^{222}$Rn activity history 
can be attributed to fluctuations in detector operating conditions, usually changes in temperature 
that generated convection currents which circulated contaminants from the balloon surface 
into the fiducial volume. 
It is worth noting that in Fig.~\ref{fig:radon-vs-time} we present the radon activity inside a 
5.5-m-radius spherical volume primarily for purpose of comparison with data from the 
previous calibration system~\cite{Busenitz:2009ac}, but we find the fundamental result is 
the same if we consider instead a smaller-radius (1.2~m) cylindrical volume centered on 
the path the MiniCal cable traveled along the detector's z-axis.

In short, we performed a variety of data-analysis studies and found no indication that 
MiniCal introduced any contamination, radon or otherwise---even during extended 
multi-day deployments---above the background levels existing in the LS following 
purification~\cite{TakemotoPhDThesis}.

\section{Conclusions}

We successfully designed, built, and operated an ultra-clean calibration system that enabled
accurate, precise positioning of radioactive sources along the z-axis of the KamLAND detector. 
The main device is simple in concept, but its realization was challenging due to the need to 
make it compact and precise in its operations while also ensuring it would not contaminate or 
degrade the detector's highly radiopure liquid scintillator.  The system was routinely used to 
deploy sources inside the detector over a period of more than two years, and no adverse 
effects on the liquid scintillator were observed. We are now in the process of modifying the 
system for use in KamLAND-Zen, an experiment utilizing the KamLAND detector to search 
for neutrinoless double-beta decay of 
$^{136}$Xe~\cite{KamLANDZen:2012aa, Gando:2012pj, Gando:2012zm}.




\section*{Acknowledgments}

We are grateful to the UC~Berkeley Physics and LBNL machine shops for
manufacturing and assembling many MiniCal components.
Thanks are also due to 
%
%
K.~Terao and C.~Shimmin for their assistance in testing MiniCal; 
Al Smith of the LBNL Low Background Facility for radioassaying materials; 
J.~Busenitz at the University of Alabama for help in ensuring that new and existing 
radioactive sources were mechanically compatible with MiniCal; 
and the former members of the full-volume calibration system team for sharing their experience and advice.
%

The KamLAND experiment is supported by the Grant-in-Aid for Specially Promoted 
Research under Grants 16002002 and 21000001 from the Japanese Ministry of Education, 
Culture, Sports, Science and Technology; the World Premier International Research 
Center Initiative (WPI Initiative), MEXT, Japan; Stichting voor Fundamenteel Onderzoek der 
Materie~(FOM) in the Netherlands; and the US Department of Energy~(DOE) under Grants 
DE-AC02-05CH11231 and DE-FG03-00ER41138, as well as other DOE grants to individual 
institutions. The Kamioka Mining and Smelting Company provided service for activities in 
the mine.


\biboptions{sort&compress}
\bibliographystyle{elsarticle-num}
\bibliography{MiniCal-NIMA-paper}

\end{document}